\documentclass[reprint, secnumarabic,amssymb, nobibnotes, aps, prl]{revtex4-1}

\usepackage{color}
\usepackage{epsfig, graphics, graphicx, subfigure, wrapfig, lipsum}
\usepackage{verbatim}
\usepackage{dcolumn}
\usepackage{bm}
\usepackage{amsmath, braket}
\usepackage{array}
\usepackage{xr}

\newcolumntype{X}[1]{>{\centering\arraybackslash\hspace{0pt}}p{#1}}
\newcolumntype{M}[1]{ >{\centering\arraybackslash}m{#1}}
\newcommand{\roml}[1]{\lowercase\expandafter{\romannumeral #1\relax}}
\newcommand{\romu}[1]{\uppercase\expandafter{\romannumeral #1\relax}}

\begin{document}

\title{Understanding Phonon Thermal Transport in Twisted Bilayer Graphene}

\author{Shahid Ahmed}
\author{Shadab Alam}
\author{Ankit Jain}
\email{a\_jain@iitb.ac.in}
\affiliation{Mechanical Engineering Department, IIT Bombay, India}
\date{\today}

\begin{abstract}
{The phonon thermal transport properties of twisted bi-layer graphene are investigated using lattice dynamics and the Boltzmann transport equation. The thermal conductivities of $13.2^{\circ}$ and $21.8^{\circ}$ twisted configurations are 56\% and 36\% lower than the untwisted configuration which has a room temperature thermal conductivity of 2260 W/m-K. The reason for this lower thermal conductivity is unraveled from phonon mode-level analysis made possible through untwisting of layers. Due to a large commensurate unitcell of twisted 
configuration, the Brillouin zone is folded but this folding has no bearing on the phonon scattering phase space. The major impact of twisting is felt by flexural phonons with out-of-plane vibrations via the change in strength of flexural interatomic interactions and since these flexural phonons carry majority of the heat in bilayer graphene (65\% at room temperature), the thermal conductivity is sensitive to layer twisting. Our study suggests that twisting will affect thermal transport only for those materials that have a major contribution from flexural phonon modes. }

\end{abstract}
\maketitle

{\bf Introduction}
Twisted bilayer graphene (TBG) is a vertically stacked arrangement of two monolayer graphene sheets where the sheets are twisted/rotated with respect to each other to form Moire patterns \cite{bistritzer2011}. Depending on the twist angle, the electronic bands undergo different degrees of hybridization and for a particular twist angle of $1.06^{\circ}$, termed as the magic angle, the electronic band hybridization in TBG results in flat bands near the Fermi level with associated massless fermions leading to  high-temperature superconductivity \cite{cao2018, cao2018a}. Since this discovery, the theoretical and experimental research interests have spurred in TBG and now this material is known to host many more exotic phenomenona such as  quantum anomalous Hall effect \cite{serlin2020,nuckolls2020,li2021,wu2021,das2021}, spontaneous orbital ferromagnetism \cite{sharpe2019,serlin2020}, topologically protected states \cite{song2019}, large spin entropy and Pomeranchuk effect \cite{rozen2021, saito2021}. 

Considering that monolayer graphene has the highest reported room-temperature thermal conductivity \cite{cai2010,balandin2011,chen2011} which originates from reduced/non-scattering of out-of-plane flexural modes \cite{lindsay2010}, the twist-angle in TBG can alter the thermal transport resulting in the different thermal conductivity of TBG compared to that of untwisted bi-layer graphene (BLG). Initial experimental findings by Li et al.~\cite{li2014} confirmed this speculation and the thermal conductivity of TBG with $34^{\circ}$ twist-angle is measured to be lower than that of BLG. This reduced thermal conductivity of TBG compared to its untwisted counter-part is hypothesized to originate from increased Umklapp phonon-phonon scattering in TBG owing to its smaller/folded Brillouin zone arising from the larger commensurate unitcell of TBG. This hypothesis is yet to be confirmed as all computational attempts so far are based on the molecular dynamics simulations \cite{li2014, li2018,liu2022,kumar2023, cheng2023} which support reduced thermal conductivity of TBG for all twist angles but the reason for this reduction largely remained elusive from these simulations. 

Further, while there is a general consensus on the reduction of thermal conductivity with twist in literature reported results, there is no agreement on the exact twist-angle dependence. For instance, Li et al.~\cite{li2018} reported around 33\% decrease in thermal conductivity with change in twist angle from $21.8^{\circ}$ to  $13.2^{\circ}$, while Cheng et al.~\cite{cheng2023} reported no change in thermal conductivity for similar variation in twist angles. Even for untwisted AB-stacked BLG, the literature reported values vary by more than 100\% from 450-700 W/m-K by Han et al.~\cite{han2021} and Liu et al.~\cite{liu2022}, 1050 W/m-K by Li et al.~\cite{li2018} to 1600 W/m-K by Kumar et al.~\cite{kumar2023}, and 2400 W/m-K by Cheng et al.~\cite{cheng2023} compared to experimentally measured values of 1900-2100 W/m-K \cite{li2014,han2021} at 300 K. In addition to other factors, this large variation in the literature reported thermal conductivity of untwisted and twisted BLG originates from large mean free paths of heat carrying phonon in BLG \cite{lindsay2011b} which renders obtained thermal conductivities to be length/size sensitive in molecular dynamics simulations \cite{liu2022}.

To this end, in this work, we study thermal transport in TBG by solving the Boltzmann transport equation (BTE) along with lattice dynamics calculations. In contrast with molecular dynamics simulations, the lattice dynamics calculations have an advantage of directly providing mode-level phonon properties along with quantum phonon statistics \cite{mcgaughey2019}. Further, the phonon properties obtained from the BTE-based approach are intrinsic phonon properties corresponding to the bulk system. 

We find that while harmonic phonon properties remain majorly unchanged in twisting graphene layers with respect to each other, the phonon-phonon scattering increases and thermal conductivity decreases. However, the origin of enhanced phonon-phonon scattering in TBG is increased anharmonicity and not the Umklapp scattering phase space due to Brillouin-zone folding as was hypothesised earlier \cite{li2014}. In fact, due to negligible change in phonon dispersion in twisted vs. untwisted bilayer graphene, the Normal and Umklapp phonon-phonon scattering phase space remains majorly unchanged with twist angle. Furthermore, we find that the origin of increased phonon anharmonicity is cross-plane interactions which affect scattering rates of flexural phonon modes. Based on our findings, we anticipate that inter-layer twisting can alter thermal transport only for those materials where monolayer transport is dominated by out-of-plane flexural phonon modes.

{\bf Methodology}
We considered two twisted TBG configurations with twist angles of $13.2^{\circ}$ and $21.8^{\circ}$  and obtained their lattice thermal conductivity by solving the BTE iteratively as \cite{reissland1973, jain2020}:
\begin{equation}
 \label{eqn_k}
    k_{ph}^{\alpha} = \sum_i c_{ph, i} v_{\alpha}^2 \tau_i^{\alpha},
\end{equation}
where the summation is over all the phonon modes in the Brillouin zone enumerated by $i\equiv(q,\nu)$, where $q$ and $\nu$ are phonon wavevector and mode index, and $c_{ph,i}$, $v_{\alpha}$, and $\tau_i^{\alpha}$ represent phonon specific heat, group velocity ($\alpha$-component), and transport lifetime respectively. As opposed to the commonly used relaxation time approximation, the iterative solution of BTE does not treat Normal phonon-phonon scattering processes as resistive and is found crucial for the correct description of the thermal transport physics in graphene \cite{lindsay2010, lindsay2014, jain2015}. The transport lifetimes are obtained by considering phonon-phonon scattering via three-phonon scattering processes and phonon-boundary scattering corresponding to a characteristic length scale, $L_{bdry}$, of 10 $\mu\text{m}$, i.e., $1/\tau_{bdry} = {L_{bdry}}/{|v|}$.
Further details regarding the calculation of thermal conductivity from the iterative solution of BTE can be found in Refs.~\cite{mcgaughey2019, jain2020}.

\begin{figure}
\centering
\epsfbox{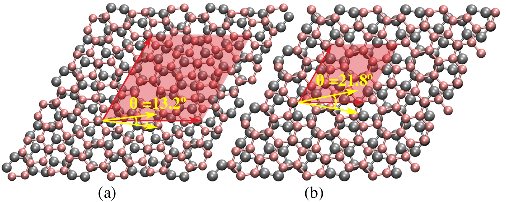}
\caption{{\bf Twisted Bilayer Graphene} The AB stacked (a) $13.2^{\circ}$ and (b) $21.8^{\circ}$ twisted bilayer graphene (TBG). The corresponding unitcells are shaded in red and have 76 and 28 atoms respectively. The atoms in top and bottom layers are represented using different colors.}
\label{fig_structure}
\end{figure}

{\bf Computational Details}
The commensurate unitcells of $13.2^{\circ}$ and $21.8^{\circ}$ TBG consists of 76 and 28 atoms respectively and are shown in Fig.~\ref{fig_structure}. The in-plane and cross-plane interatomic interactions between carbon atoms in TBG are modelled using the Tersoff and Lennard-Jones potentials respectively. For Tersoff potential, we use the parameters reported by Lindsay et al.~\cite{lindsay2010b} which are optimized for thermal properties of carbon systems and for Lennard-Jones we use parameters reported by Liu et al.~\cite{liu2022} which results in inter-layer spacing of $3.423$ $\text{\AA}$ for AB stacked untwisted BLG.  In the extraction of interatomic force constants, the in-plane interaction cutoff is set at $2^{nd}$-neighbor shell and the cross-plane interactions are included up to $5.0$ $\text{\AA}$. The force constants are extracted using the finite difference approach using displacement size of $0.001$ $\text{\AA}$ \cite{alam2023}. The phonon-phonon scattering rates are obtained by considering three-phonon scattering using phonon wavevector grids of size $16\times16\times1$ and $24\times24\times1$ for $13.2^{\circ}$ and $21.8^{\circ}$ TBG. { The four-phonon scattering is not included in this study as this higher-order scattering is shown to bring less than 10\% change in the predicted thermal conductivity of untwisted bilayer graphene at an expense of orders of magnitude higher computational cost\cite{alam2023}. }
All reported thermal properties are at a temperature of 300 K using TBG thickness of $6.8$ $\text{\AA}$.

{\bf Results}

\begin{figure}
\centering
\epsfbox{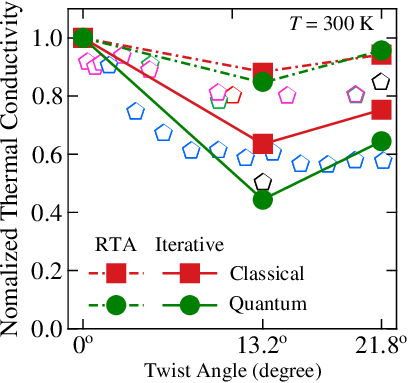}
\caption{{\bf Thermal Conductivity} The variation of normalized thermal conductivity of TBG with twist angle. The thermal conductivities are normalized using values of corresponding untwisted configurations. Results obtained using  the relaxation time approximation (RTA) and iterative/full of BTE are shown using dashed and solid lines. {The pentagonal data-points are literature reported results taken from: Refs.~\cite{liu2022} (blue: $x$-direction), \cite{cheng2023} (pink), \cite{li2018} (black), \cite{kumar2023} (green), and \cite{han2021} (red).}}
\label{fig_normalizedThermalK}
\end{figure}

Our main result for the thermal conductivity of TBG as a function of twist angle is reported in Fig.~\ref{fig_normalizedThermalK}. For comparison with the literature-reported molecular dynamics based results, along with quantum Bose-Einstein statistics, we also report thermal conductivities obtained from the classical Boltzmann statistics in Fig.~\ref{fig_normalizedThermalK}.

Using quantum statistics and iterative/full solution of the BTE, our predicted thermal conductivity at 300 K is 2260 W/m-K for untwisted BLG which is in direct agreement with experimental measurements \cite{li2014,han2021}. With twisting, we find that the thermal conductivity reduces by 56\% and 36\% for twist angles of $13.2^{\circ}$ and $21.8^{\circ}$ TBG. To confirm that these reductions are not simulation artefacts arising from the choice of the computational cell, we also evaluated thermal conductivity of untwisted BLG by starting with $13.2^{\circ}$ and $21.8^{\circ}$ structures and untwisting them to obtain untwisted BLG configurations. We find that the thermal conductivities obtained using these large unitcells (consisting of 76 and 28 atoms) are within 12\% of those obtained using the primitive 4-atoms unitcell. In particular, for untwisted and $13.2^{\circ}$-twisted configurations, we obtained thermal conductivities of 2330 and 1000 W/m-K using the same 76 atoms unitcells compared to 2260 and 2330 W/m-K for untwisted configurations using 4 atom and 76 atom (untwisted-$13.2^{\circ}$) unitcells; thus confirming that the obtained reduction in thermal conductivity of twisted configuration is not arising from the large commensurate unitcell of TBG.

\begin{figure}
\centering
\epsfbox{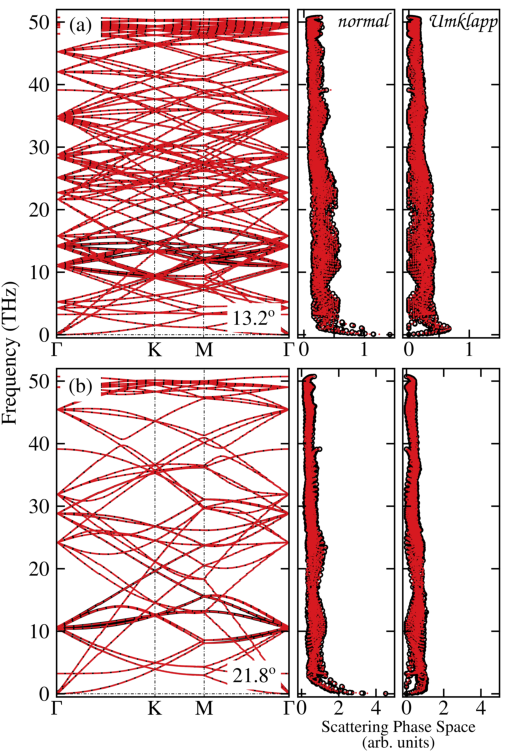}
\caption{{\bf Phonon Dispersion and Scattering Phase Space} The phonon dispersions (left panel) and normal and Umklapp three-phonon scattering phase (right panel) for (a) $13.2^{\circ}$ and (b) $21.8^{\circ}$ twisted TBG (red color). The untwisted BLG  phonon modes obtained using the corresponding untwisted configurations are included as a common reference (in black color) in (a) and (b).}
\label{fig_dispersion}
\end{figure}

To gain insights into the reduction of thermal conductivity with layer twisting, we investigate phonon dispersion and three-phonon scattering phase space in Fig.~\ref{fig_dispersion}. 
We also include phonon properties for untwisted BLG by employing larger unitcells obtained from untwisting of $13.2^{\circ}$ and $21.8^{\circ}$ TBG.  The properties obtained from these untwisted BLG are shown in black in Fig.~\ref{fig_dispersion} and enables mode-level comparison of phonon modes between twisted/un-twisted configurations. 

We find that, similar to findings of Cocemasov et al.~\cite{cocemasov2013} and Li et al.~\cite{li2018}, while there are more phonon branches in $13.2^{\circ}$ TBG arising from larger commensurate unitcell, the phonon dispersions  remains largely unchanged with twisting of layers. Consequently, the normal and Umklapp phonon-phonon scattering phase spaces, which are obtained by counting the fraction of total three-phonon processes capable of satisfying energy and crystal momentum conservation selection rules [without and with $G$ (reciprocal lattice vector) for normal and Umklapp processes], also remains unchanged between twisted and untwisted configurations.
This suggests that (\roml{1}) the phonon mode-dependent heat capacity and group velocities are same for twisted and untwisted configurations and, therefore, according to Eqn.~\ref{eqn_k}, the reduction in thermal conductivity of TBG as compared to BLG is due to a reduction in phonon transport lifetimes and (\roml{2}) since the phonon-phonon scattering phase space remains unchanged between twisted and untwisted configurations, the reduction in phonon lifetimes could only come from changes in phonon anharmonicity arising from anharmonic/cubic force constants.

\begin{figure}
\centering
\epsfbox{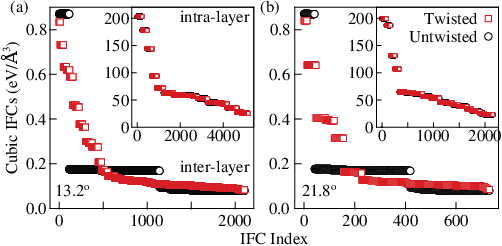}
\caption{{\bf Inter-atomic Force Constants (IFCs)} The inter-layer (main figure) and intra-layer (insets) cubic IFCs for (a) $13.2^{\circ}$ and (b) $21.8^{\circ}$ TBG. Only inter-layer IFCs undergo changes with layer twisting. The inter-layer IFCs are two orders of magnitude smaller than the intra-layer IFCs.}
\label{fig_IFCs}
\end{figure}

\begin{figure}
\centering
\epsfbox{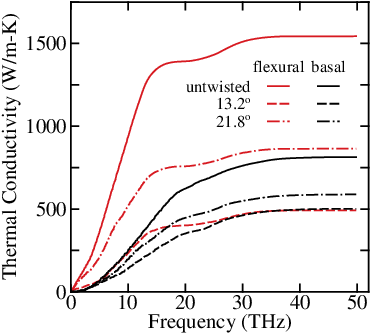}
\caption{{\bf Thermal Conductivity Accumulation} The contribution of flexural (red) and basal-plane (black) dominated phonon modes towards the thermal conductivity of untwisted and twisted bi-layer graphene. The modes are identified as flexural- vs basal-dominated based on eigen-displacements of atoms. The contribution of flexural modes decreases by more than three-fold from 1570 W/m-K for untwisted BLG to 500 W/m-K in $13.2^{\circ}$ TBG.}
\label{fig_accumulation}
\end{figure}

The cubic interatomic force constants are compared for different layer twists in Fig.~\ref{fig_IFCs}. The intra-layer interactions arising from atomic interactions within the same layer are unchanged between untwisted and twisted configurations [insets of Figs.~\ref{fig_IFCs}(a) and \ref{fig_IFCs}(b) for $13.2^{\circ}$ and $21.8^{\circ}$ twists]. The inter-layer force constants, i.e., involving atoms from both layers, are strongly dependent on layer twist and are different for untwisted and $13.2^{\circ}$- and $21.8^{\circ}$-twisted TBG. The strength of these inter-layer force constants is, however, two orders of magnitude weaker than those of the intra-layer force constants. As such, it is difficult to comprehend how the changes in strength of inter-layer cubic force constants could bring up to a factor of two difference in the thermal conductivity of TBG. To understand this, we calculate the contribution of different phonon modes towards the thermal conductivity and report it separately for flexural- and basal-plane dominated modes in Fig.~\ref{fig_accumulation}. We classify phonon mode as a flexural/basal-mode depending on its eigen-displacement, i.e., a mode is classified as flexural when $|e_{i,flexural}^{2}|$ $>$ $|e_{i,basal}^{2}|$, where $e_{i,flexural}$, $e_{i,basal}$ are flexural- and basal-plane components of eigenvector, $e_i$, of mode $i$. 

We find that in untwisted BLG, flexural modes contribute 1570 W/m-K (65\% of total) to the thermal conductivity. This flexural contribution is strongly affected by layer twisting and with $13.2^{\circ}$ and $21.8^{\circ}$ twists, flexural modes contribution reduces to only 500 and 880 W/m-K. In contrast, the basal modes are less affected and their contribution reduces from 825 W/m-K for untwisted BLG to 510 and 600 W/m-K in $13.2^{\circ}$ and $21.8^{\circ}$ TBG. This suggests that the reduction in thermal conductivity of TBG is predominantly a consequence of reduced flexural modes contribution arising from enhanced phonon-phonon scattering due to a larger spread of flexural cubic force constants in TBG. 

{\bf Discussion}
Since the discovery of magic angle graphene, there has been an ever-growing research interest in various properties of twisted layered materials. For thermal transport, the widely studied system is TBG for which experimentally it is confirmed that the thermal conductivity is 20-25\% lower than that of BLG depending on the twist angle. The origin of this reduction is speculated to be enhanced Umklapp phonon-phonon scattering arising from larger commensurate unitcell of TBG. 

In this work, we found that in agreement with experiments and literature studies, the thermal conductivity of TBG is indeed lower than that of BLG. However, this reduction in thermal conductivity of TBG is not due to a change in the scattering phase of phonon-phonon scattering arising from the reduced Brillouin zone as speculated in the literature. Rather, we find that the thermal conductivity of TBG is low due to reduced contribution of flexural phonon modes arising from enhanced phonon-phonon scattering owing to change in anharmonicity of flexural modes. The anharmonicity of these flexural modes is sensitive to layer twisting and the contribution of these modes reduces by more than three-fold to 500 W/m-K in $13.2^{\circ}$ TBG compared to more than 1500 W/m-K in untwisted BLG.

Our finding has an important consequence on the thermal transport properties of twisted materials: our study suggests that, with twisting, while there will be a Brillouin zone folding in all materials, the thermal transport will be affected only for those materials which has a large contribution from flexural modes.

{ Finally, it is worth mentioning that in this study, the intra-layer interactions are unchanged with twisting of layers while in experiments, these interactions could vary. In fact, the strong electronic correlations arising from layer twisting close to the magic angle is responsible for novel opto-electronic properties of TBG. While these changes in interactions are not accounted for in this study, they will be considered in our future studies. }

{\bf Conclusions}
In summary, we investigated the thermal transport properties of twisted bilayer graphene using iterative solution of the Boltzmann transport equation. We find that the phonon dispersion and phonon-phonon scattering phase space are unchanged with twisting of layers. The thermal conductivities of twisted layers are lower than that of untwisted layers and this reduction in thermal conductivity arises from flexural phonons which are sensitive to layer twisting and contribute more than 65\% to the thermal conductivity of untwisted bilayer graphene. 

Based on our findings, we conclude that due to a large commensurate unitcell even though twisting results in Brillouin zone folding for all twisted materials, the extent of thermal conductivity change with twisting is governed predominantly by the contribution of flexural phonon modes towards thermal transport in untwisted layers.

{\bf Acknowledgement}
The authors acknowledge the financial support from National Supercomputing Mission, Government of India (Grant Number: DST/NSM/R\&D-HPC-Applications/2021/10) and Core Research Grant, Science \& Engineering Research Board, India (Grant Number: CRG/2021/000010).  The calculations are carried out on SpaceTime-II supercomputing facility of IIT Bombay and PARAM Sanganak supercomputing facility of IIT Kanpur.

{\bf Data Availability}
The raw/processed data required to reproduce these findings is available on a reasonable request via email.

\bibliographystyle{plain}

\end{document}